\documentclass[aps,nofootinbib,twocolumn,longbibliography]{revtex4-1}
\usepackage{graphicx,amsmath,amssymb,amsbsy,latexsym,verbatim}
\usepackage{graphicx}

\usepackage{hyperref}

\newcommand{\be}{\nopagebreak[3]\begin{equation}}
\newcommand{\ee}{\end{equation}}
\newcommand{\ba}{\nopagebreak[3]\begin{eqnarray}}
\newcommand{\ea}{\end{eqnarray}}

\newcommand{\bc}{\begin{comment}}
\newcommand{\ec}{\end{comment}}

\begin{document}
\title{A note on the geometrical interpretation of \\ quantum groups and non-commutative spaces  in  gravity}
 \author{Eugenio Bianchi}
 \author{Carlo Rovelli}
    \affiliation{Centre de Physique Th\'eorique de Luminy \footnote{Unit\'e mixte de recherche du CNRS et des Universit\'es de Aix-Marseille I, II et du Sud; affili\'e \`a la FRUMAM.}, Case 907, F-13288 Marseille, EU}
\begin{abstract} 

\noindent  Quantum groups and non-commutative spaces have been repeatedly utilized  in approaches to quantum gravity.  They  provide a mathematically elegant cut-off,  often interpreted as related to the Planck-scale quantum uncertainty in position.   We consider here a different geometrical interpretation of this cut-off, where the relevant non-commutative space is the space of \emph{directions} around any spacetime point. The  limitations in angular resolution expresses the finiteness of the angular size of a Planck-scale minimal surface at a maximum distance $1/{\sqrt{\Lambda}}$ related the cosmological constant $\Lambda$.  This yields a simple geometrical interpretation for the relation between the quantum deformation parameter $q=e^{i\Lambda\,l^2_{\rm Planck}}$  and the cosmological constant, and resolves a difficulty of more conventional interpretations of the physical geometry described by quantum groups or fuzzy spaces.
\end{abstract}
\date{ \today}

\maketitle

If we look at a small sphere of radius $l$ which is at a large distance $L$, we see it under an angle $\phi\sim l/L$.  If we live in a 3-sphere with large radius $L_{max}\equiv 1/{\sqrt{\Lambda}}$, or in a Lorentzian space with a horizon at that distance, we will never see the sphere under an angle  smaller than ${l}/L_{max}$.  Suppose in addition that for some reason no object with size smaller than $l_{min}\equiv l_P$  exists in the universe.  See Figure 1.   Then we will never see anything having angular size smaller than  
\be
\phi_{min}\sim \frac{l_{min}}{L_{max}}=\sqrt{\Lambda}\, l_P.  
\label{eq:min angle}
\ee 
In such a situation, everything we see is captured, on the local celestial 2-sphere formed by the directions around us, by spherical harmonics with $j \le j_{max}$. Since the  $j$-th spherical harmonic distinguishes dihedral angles of size $\phi^2\sim4\pi/(2j+1)$, we won't see harmonics with 
\be
j>j_{max}=4\pi/\phi_{min}^2\sim \frac{4\pi}{l^2_P\, \Lambda}.
\ee  

A 2-sphere not resolved at small angles is a ``fuzzy sphere" \cite{Madore:1991bw,Madore:1997ta,Madore:2002fk,Freidel:2001kb}, and is described by the algebra of the angular functions spanned by the spherical harmonics with   $j\le j_{max}$. 
 Alternatively, a maximum angular momentum characterizes the representations of the  
quantum group $SU_q(2)$, when $q=e^{i2\pi/k}$ and $k\sim 2j_{max}$  \cite{Majid:1988we,Maggiore:1993zu,Majid:2000fk}. 

 It follows that in a  universe characterized by a maximum visibility $L=1/{\sqrt{\Lambda}}$ and a minimal length $l_P$, the local rotational symmetry is better described by $SU_q(2)$ than by $SU(2)$, with
\be
                 q=e^{i\Lambda l^2_P}.
                 \label{q}
\ee 

\begin{figure}[h]
\centerline{\includegraphics[scale=0.15]{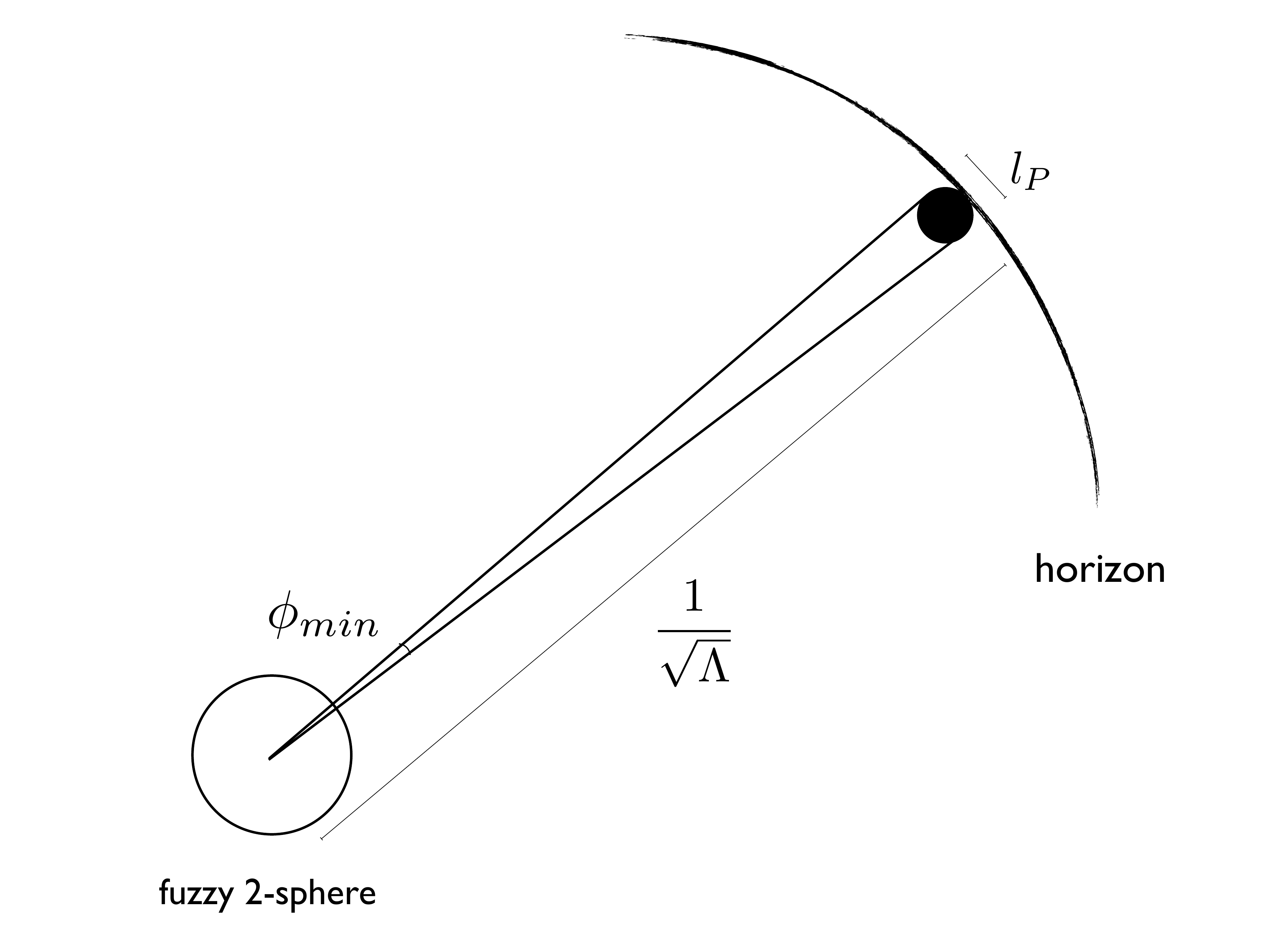}}
\caption{{The fuzzy sphere is the local celestial sphere.}} 
\vspace{-2mm}
\end{figure}

Equation \eqref{q} is precisely the relation between the cosmological constant $\Lambda$, the Planck length $l_P$ and the $SU_q(2)$ quantum deformation parameter $q$ that emerges from  quantizations of general relativity with a cosmological constant (see for instance \cite{Major:1995yz,Smolin:2002sz,Freidel:1998pt}).  

A minimal Planck-scale area  $A_o\sim l^2_P$ for any observed physical object is actually predicted by the quantization of general relativity in several approaches to quantum gravity and in particular in loop gravity \cite{Rovelli:1994ge} (See \cite{Rovelli:2011eq} for a recent introduction). Similar indications of a minimal observable scale come also from string theory \cite{AmatiVeneziano3}. A maximal distance of the order $a\sim 1/\sqrt{\Lambda}$ is also essentially implied by general relativity, when a cosmological constant  $\Lambda$ is present.  Therefore the combination of the quantum gravitational space granularity with the maximal size determined by the cosmological constant yields immediately a local quantum-group structure with a deformation parameter given by \eqref{q}.

If this is correct, there should be a physical fuzziness -- described by the quantum group $SU_q(2)$, or by a fuzzy sphere -- of the local ``celestial sphere" seen by any observer in a universe where physics is characterized by a minimal and a maximal scale. 

A similar conclusion can be reached from the manner in which a quantum group appears in the quantizations of general relativity with cosmological constant.  The quantum deformation affects the local gauge.  The local gauge  of general relativity in the time gauge is $SU(2)$, and its physical interpretation is as the  covering of the group of  physical rotations around any given point in space. A quantum deformation of this  group corresponds, physically, to a non-commutativity, and therefore to a consequent intrinsic fuzziness, of any angular function. In other words, it describes the impossibility of resolving small dihedral angles of view. 

This picture is realized concretely in loop gravity, where the angle $\phi$ between two directions in space is an operator with a discrete spectrum  \cite{Penrose2,Major:1999mc}. Its eigenvalues are labeled by two spins $j_1$, $j_2$, associated to the two directions, and a quantum number $k=|j_1-j_2|,\ldots, j_1+j_2$. They are given by 
\be
\cos \phi=\frac{k(k+1)-j_1(j_1+1)-j_2(j_2+1)}{2\sqrt{j_1(j_1+1)}\sqrt{j_2(j_2+1)}}\;.
\label{eq:angle}
\ee
If the spins are bounded by $j_{max}\gg 1$, then the best angular resolution is 
\be
\phi_{min}=\sqrt{2/j_{max}}\;.
\ee
In the presence of a cosmological constant $\Lambda$, the bound is $j_{max}\sim (l_P^2\, \Lambda)^{-1}$, and  relation (\ref{eq:min angle}) is reproduced.

Quantum groups and non-commutative spaces have been repeatedly utilized in approaches to quantum gravity \cite{Snyder:1946qz,Chamseddine:1992yx,Madore:1991bw,Madore:1997ta,Madore:2002fk,Jevicki:2000it,Moffat:2000gr,Vacaru:2000yk,Cacciatori:2002ib,Cardella:2002pb,Vassilevich:2004ym,Buric:2006di,Abe:2002in,Valtancoli:2003ve,Kurkcuoglu:2006iw,Krajewski:1999bg,Grosse:2004yu,Freidel:2005me,Szabo:2006wx,Majid:1988we,Maggiore:1993zu,Majid:2000fk,Livine:2008hz}. The associated non-commutative spaces are generally interpreted as describing Planck-scale quantum uncertainty in position.  The geometrical picture suggested here is  different: fuzziness is in the directions, not in the distances. This may resonate with the spectral point of view on space implicit in \cite{albook} and \cite{AmelinoCamelia:2011bm} and with the 
4d-angle (speed) quantization in \cite{Girelli:2003az}. 

This geometrical picture resolves also a certain difficulty in interpreting the non-commutativity implied by a cut-off in the spins as related to the Planck scale fuzziness of physical space: the deformation parameter is dimensionless, and by itself does not determine a scale at which physical space becomes fuzzy.  Introducing $\hbar$ does not help, since $\hbar$ contains a mass.  The quantization of physical space is a direct quantum-gravitational effect, which has probably no direct relation with quantum groups, and is governed by $\hbar$ and the Newton constant  \cite{Rovelli:1994ge}.  In order for a quantum group to play a geometrical role, a second dimensional quantity is needed, and this is provided by a maximal distance in space, as implied by the presence of a cosmological constant. 

The geometry of the covariant picture \cite{E.Buffenoir:kx,Noui:2002ag,Han:2010pz,Fairbairn:2010cp,Wieland:2011kx,Han:2011vn} will be discussed elsewhere.

\medskip

\bibliographystyle{apsrev4-1}
\bibliography{BiblioCarlo}

\end{document}